# Nature of the metal-insulator transition in few-unit-cell-thick LaNiO$_3$ films

M. Golalikhani[1], Q. Lei[1], R.U. Chandrasena[1,2], L. Kasaei[1], H. Park[3,4], J. Bai[5], P. Orgiani [6,7], J. Ciston[8], G.E. Sterbinsky[9], D.A. Arena[10], P. Shafer[11], E. Arenholz[11], B.A. Davidson[1,7], A.J. Millis[12,13], A.X. Gray[1,2] & X.X. Xi[1,2]

The nature of the metal-insulator transition in thin films and superlattices of LaNiO$_3$ only a few unit cells in thickness remains elusive despite tremendous effort. Quantum confinement and epitaxial strain have been evoked as the mechanisms, although other factors such as growth-induced disorder, cation non-stoichiometry, oxygen vacancies, and substrate–film interface quality may also affect the observable properties of ultrathin films. Here we report results obtained for near-ideal LaNiO$_3$ films with different thicknesses and terminations grown by atomic layer-by-layer laser molecular beam epitaxy on LaAlO$_3$ substrates. We find that the room-temperature metallic behavior persists until the film thickness is reduced to an unprecedentedly small 1.5 unit cells (NiO$_2$ termination). Electronic structure measurements using X-ray absorption spectroscopy and first-principles calculation suggest that oxygen vacancies existing in the films also contribute to the metal-insulator transition.

[1] Department of Physics, Temple University, Philadelphia, PA 19122, USA. [2] Temple Materials Institute, Temple University, Philadelphia, PA 19122, USA. [3] Department of Physics, University of Illinois at Chicago, Chicago, IL 60607, USA. [4] Materials Science Division, Argonne National Laboratory, Argonne, IL 60439, USA. [5] National Synchrotron Light Source II, Brookhaven National Laboratory, Upton, NY 11973, USA. [6] CNR-SPIN, UOS Salerno, 84084 Fisciano, Italy. [7] CNR-IOM, TASC Laboratory in Area Science Park, 34139 Trieste, Italy. [8] National Center for Electron Microscopy Facility, Molecular Foundry, Lawrence Berkeley National Laboratory, Berkeley, CA 94720, USA. [9] Advanced Photon Source, Argonne National Laboratory, Argonne, IL 60439, USA. [10] Department of Physics, University of South Florida, Tampa, FL 33620, USA. [11] Advanced Light Source, Lawrence Berkeley National Laboratory, Berkeley, CA 94720, USA. [12] Department of Physics, Columbia University, New York, NY 10027, USA. [13] Center for Computational Quantum Physics, The Flatiron Institute, New York, NY 10010, USA. Correspondence and requests for materials should be addressed to M.G. (email: maryam@temple.edu) or to X.X.X. (email: xiaoxing@temple.edu)

The metal-insulator transition (MIT) in LaNiO$_3$ ultrathin films and superlattices[1–6] has been proposed to be affected by reduced dimensionaltiy and epitaxial strain, raising the possibility of controlling electronic phases of strong electron correlation materials by nanoscale engineering. In bulk form, LaNiO$_3$ is the only member of the perovskite series $R$NiO$_3$ ($R$: rare earth) that is a paramagnetic metal not exhibiting an MIT at any temperature[7]. In the form of a thin layer, either as a thin film[8–10] or as a part of a superlattice[3,11], it shows insulating behavior (d$\rho$/d$T$ < 0) at room temperature when its thickness is reduced below a few unit cells (u.c.), 2 u.c. being the lowest critical thickness value reported prior to this work[4]. Numerous prior studies attributed this phenomenon to dimensional crossover[4,10] perhaps accompanied by structural distortions[5,12,13]. It has been suggested that the orbital order in LaNiO$_3$ can be tailored via a combination of reduced dimensionality and epitaxial strain to resemble that of high-temperature cuprate superconductors[14,15]. Large orbital polarization has been experimentally observed in LaNiO$_3$ superlattices[16].

In this work, our first-principles calculations show that dimensionality and strain alone are not sufficient to induce the MIT in LaNiO$_3$ on a LaAlO$_3$ substrate even at one-unit-cell thickness. To reveal the nature of the transition, we use atomic layer-by-layer laser molecular beam epitaxy (ALL-Laser MBE)[17] from two separate binary oxide targets (La$_2$O$_3$ and NiO) for LaNiO$_3$ film growth, which allows building up the film one atomic layer at a time under a high oxygen pressure. We find that the LaNiO$_3$ films thus produced remain metallic at room temperature down to a thickness of 1.5 u.c. (NiO$_2$ termination). Electronic structure measurements using X-ray absorption spectroscopy (XAS) show a feature in the spectra that has been attributed to apical oxygen vacancies. We conclude that in addition to low dimensionality and strain, oxygen vacancies play an important role in driving the ultrathin LaNiO$_3$ films to become insulating.

## Results

**Atomic layer-by-layer growth.** Figure 1a illustrates the atomic layer-by-layer synthesis of LaNiO$_3$ films on LaAlO$_3$ substrates. Reflection high-energy electron diffraction (RHEED) intensity oscillation patterns during growth are shown for a series of LaNiO$_3$ films from 0.5 u.c. to 3.5 u.c. in thickness. The half-integer u.c. denotes NiO$_2$-terminated films while the integer numbers correspond to LaO-terminated films. On a LaAlO$_3$ substrate, a homoepitaxial LaAlO$_3$ buffer layer was first grown, achieved in situ by alternately ablating La$_2$O$_3$ and Al$_2$O$_3$ targets, until the RHEED intensity pattern confirmed a highly crystalline surface. This step ensures the near-ideal epitaxial growth of LaNiO$_3$ starting with the first NiO$_2$ atomic layer (the first half unit cell). Deposition of different number of NiO$_2$ and LaO layers leads to LaNiO$_3$ films with different number of unit cells and termination.

The ultrathin LaNiO$_3$ films were further characterized by X-ray reflectivity (XRR) and scanning transmission electron microscopy (STEM). Figure 1b shows a momentum-dependent XRR spectrum for the 2 u.c. LaNiO$_3$ film on a 5 u.c. LaAlO$_3$ buffer layer (open circles) as well as the result of an IMD simulation (solid line) taking into account the roughness at both the surface and interface. The best fit to the spectrum, including the minima positions and the shape of the oscillation, indicates a thickness of 8.0 ± 1.5 Å for the LaNiO$_3$ film and 20.0 ± 2.0 Å for the LaAlO$_3$ buffer layer, in agreement with the growth control parameters.

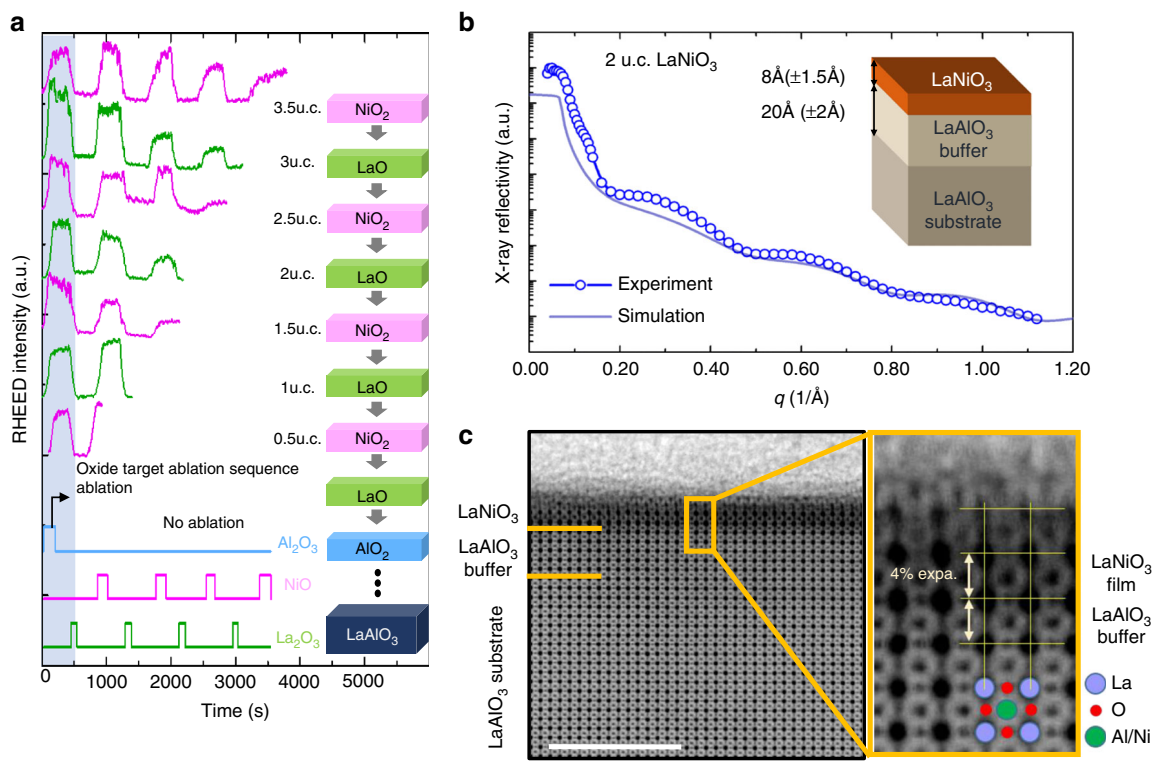

**Fig. 1** Growth and surface characterization of ultrathin LaNiO$_3$ films. **a** RHEED intensity oscillation during the growth of the homoepitaxial LaAlO$_3$ buffer layer and 0.5 to 3.5 u.c. of LaNiO$_3$ ultrathin films with both NiO$_2$ and LaO surface terminations. **b** X-ray reflectometry measurements for a 2 u.c. LaNiO$_3$ film with LaO surface termination. **c** ABF STEM image of the LaNiO$_3$/LaAlO$_3$ interface in cross-sectional geometry with black atom contrast and drift correction applied. Scale bar, 5 nm. Also shown is the magnification of the interface area with schematic atomic structure overlay identifying both metal and oxygen sites. The LaNiO$_3$ layers exhibited an out of plane expansion of 4 ± 0.5% as compared to the LAO substrate with no measurable in-plane expansion

The drift-corrected[18] Annular Bright Field (ABF) STEM image of the same sample, obtained in the cross-sectional geometry with black atom contrast, is shown in Fig. 1c. The LaNiO$_3$ layers appear darker due to their higher average Z number. By counting the atomic layers of the LaNiO$_3$ film, we determine that the film thickness is between 2 and 2.5 u.c. The 5 u.c. LaAlO$_3$ homoepitaxial buffer layer is virtually indistinguishable from the underlying LaAlO$_3$ substrate in the STEM image, indicative of its high crystallinity. From the image the out-of-plane lattice constant of the LaNiO$_3$ unit cell is measured to be 3.94 Å, 4 ± 0.5% larger than the in-plane lattice constant. This may represent a volume expansion, which has been attributed previously to oxygen vacancies[19,20], although NiO$_6$ octahedral rotations are also known to affect bond angles and lengths in LaNiO$_3$ thin films[21]. X-ray diffraction reciprocal space map of a 50-u.c.-thick film shows a fully strained film, suggesting that all the LaNiO$_3$ films presented in the work are fully compressively strained to the LaAlO$_3$ substrate.

**Metal-insulator transition.** Resistivity vs. temperature curves for LaNiO$_3$ films with thicknesses ranging from 1 to 4.5 u.c. are presented in Fig. 2a along with data from a 50 u.c. film. We first focus on the room-temperature resistivity, $\rho$(300 K), which we classify as metallic or insulating according to the sign of the temperature derivative of the resistivity $d\rho/dT$ (positive = metallic; negative = insulating). Our LaNiO$_3$ films show metallic behavior at room temperature down to 2 u.c. We also define the MIT temperature, $T_{MIT}$, as the temperature at which $d\rho/dT = 0$. Films 3.5 u.c and thicker remain metallic down to the lowest temperature, while thinner films display an MIT in our temperature range, except for the 1 and 1.5 u.c. films which remain insulating up to the highest temperature studied. The critical film thickness for the MIT is 0.5 u.c. less than the smallest value reported previously by King et al.[4] We believe the difference from previous results arises from the use of homoepitaxial LaAlO$_3$ buffer layer that reduces defects and from improved oxygenation.

The dependence of $T_{MIT}$ and $\rho$(300 K) on film thickness and surface termination is shown in Fig. 2b, c, respectively. In general, both parameters increase with decreasing LaNiO$_3$ film thickness except that the MIT temperature and resistivity of the 2 u.c. film are lower than the 2.5 u.c. film. Kumah et al. also reported differences between integer and half-integer thicknesses (i.e. NiO$_2$ vs. LaO terminations) albeit at larger film thicknesses[5]. These authors suggest that an ionic buckling in the near-surface NiO$_2$

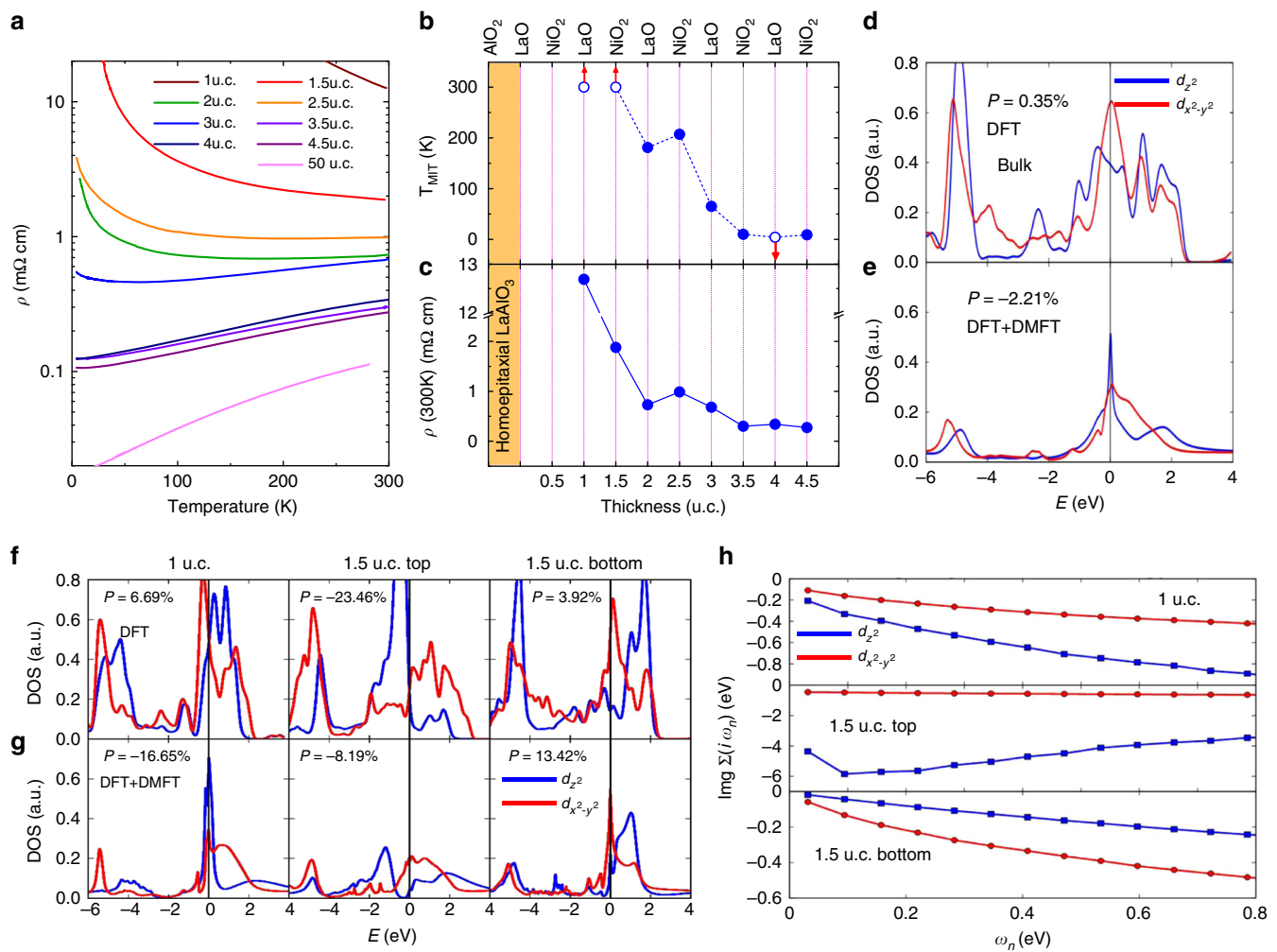

**Fig. 2** Metal-insulator transition in ultrathin LaNiO$_3$ films. **a** Temperature-dependent resistivity ($\rho$) of LaNiO$_3$ films of different thicknesses. **b** Metal-insulator transition temperature $T_{MIT}$ and **c** $\rho$(300 K) of LaNiO$_3$ films with different thicknesses and surface terminations. Open circles indicate that no MIT was observed in the temperature range studied. **d** DFT and **e** DFT + DMFT calculations of the DOS of the Ni 3d orbitals for bulk LaNiO$_3$ on the compressive LaAlO$_3$ substrate. **f** DFT and **g** DFT + DMFT calculations of the DOS for 1 u.c. and 1.5 u.c. LaNiO$_3$ thin films on the compressive LaAlO$_3$ substrate. The orbital polarization $P$ for each case is indicated. **h** The imaginary part of the DMFT self-energies of Ni $e_g$ orbitals on the Matsubara frequency for 1 u.c. and 1.5 u.c LaNiO$_3$ thin films

planes occurs for films terminated with the negatively charged NiO$_2$ plane while they are bulk-like when the films are terminated with the positively charged LaO plane[5]. We argue here that charge transfer is also important.

**Electronic structure calculations**. We performed both density functional theory (DFT) and density functional theory plus dynamical mean field theory (DFT + DMFT) calculations of the density of state (DOS) of the Ni 3$d$ orbitals for 1 u.c. (LaO termination) and 1.5 u.c. (NiO$_2$ termination) LaNiO$_3$ films on LaAlO$_3$ substrates. For comparison, the DFT and DFT + DMFT results for bulk LaNiO$_3$ with lattice parameters corresponding to the LaAlO$_3$ substrate-induced strain are shown in Fig. 2d, e, respectively. Just as the bulk LaNiO$_3$ is metallic, both 1 u.c. and 1.5 u.c. films are metallic as well within both DFT (Fig. 2f) and DFT + DMFT (Fig. 2g): the calculations exhibit nonzero spectral weights of Ni $e_g$ orbitals at the Fermi energy. However, the metallic state of the 1 u.c. film is marked by a very narrow bandwidth (mass renormalization of about a factor of 4–5) of the $3z^2-r^2$ orbital while the $x^2-y^2$ band is effectively two dimensional; these dimensionality and correlation effects suggest that the metallic states are fragile; easily disrupted by disorder, oxygen vacancies, or physics beyond the single-site DMFT approximation used here.

The calculations also reveal the importance of interlayer charge transfer and indicate interesting thickness-dependent electronic phases. We focus first on the 1 u.c. results. At the DFT level, the 1 u.c. LaNiO$_3$ thin film shows enhanced $d_{3z^2-r^2}$ hole density relative to $d_{x^2-y^2}$. This positive orbital polarization $P$, defined as

$$P = \frac{n_{x^2-y^2} - n_{3z^2-r^2}}{n_{x^2-y^2} + n_{3z^2-r^2}}, \quad (1)$$

where $n_{x^2-y^2}$ and $n_{3z^2-r^2}$ represent the number of electrons in orbitals with $x^2-y^2$ and $3z^2-r^2$ symmetries, respectively, is due to the strong quantum confinement effect which decreases the bandwidth of the $3z^2-r^2$ orbital relative to that of the $x^2-y^2$ orbital, overcoming the level shift (pushing the $3z^2-r^2$ down) from the strain effect and leading to positive orbital polarization. The strong electronic correlations treated in DFT + DMFT reduce the electron kinetic energy, decreasing the effect of quantum confinement. Orbitial polarization is then driven by the level splitting, and has the opposite sign to the prediction of the DFT calculation. The spectra calculated for the 1 u.c. LaNiO$_3$ film are consistent with those of the interfacial layer in (LaNiO$_3$)$_4$/(LaAlO$_3$)$_4$ superlattices computed previously also using DFT +DMFT[22].

The electronic structure of the 1.5 u.c. film is more interesting, distinct from either the 1 u.c. layer or bulk. The top layer of the 1.5 u.c. film is terminated as the NiO$_2$ surface and each Ni ion is only five-fold coordinated, with the vacuum interface missing the apical oxygen ion. The missing ion adds charge to the surface bands leading to strong charge transfer and to new physics. We find that the surface layer is an orbital-selective Mott insulator, with the $d_{3z^2-r^2}$ orbitals in a Mott insulating state (see the large self-energy shown in Fig. 2h) while the other band is partly filled and incoherent, with carriers strongly scattered by the orbitally selective Mott degrees of freedom (see the suppression of the quasiparticle peak and the large but weakly frequency-dependent self-energy). The bottom layer of the 1.5 u.c. film has six-fold coordinated Ni ions, but is depleted with respect to the top layer. We see a nearly empty $3z^2-r^2$ quasiparticle band (note the very small self-energy) and an effectively two dimensional $x^2-y^2$ orbital which may again be susceptible to the localization.

**X-ray absorption spectroscopy**. Detection of oxygen vacancies in thin films is challenging; however, advanced X-ray spectroscopic techniques probing electronic states that can be modified by oxygen vacancies may provide clues[23]. Figure 3a shows the polarization-averaged XAS results of LaNiO$_3$ films of various thicknesses at the O K edge, which probes the O 2$p$-projected unoccupied DOSs resulting from dipole-allowed X-ray transitions from the 1$s$ core shell[23]. The XAS spectra are normalized to zero and unity at the O K pre-edge and post-edge, respectively, and the spectra are shifted vertically for clarity. The pre-edge peak centered at approximately 528 eV (shown in the shaded box in Fig. 3a and normalized in Fig. 3b) reflects the states at the bottom of the unoccupied conduction band via O 2$p$–Ni 3$d$ orbital hybridization[24–26]. The intensity of the peak is directly related to the hole density, which in the absence of O vacancies depends on the degree of Ni–O hybridization and a higher intensity corresponds to a more metallic LaNiO$_3$ layer[26]. As seen in the figure, although the intensity of the Ni 4$sp$ feature is relatively constant for all the film thicknesses, the pre-edge peak intensity decreases as the film thickness is reduced, with the reduction being relatively more substantial for the 1.5 and 2 u.c. films. This is consistent with the substantially higher resistivity in the 1.5 and 2 u.c. films (see Fig. 2a, c). The reduction in pre-peak intensity has been attributed to reduced Ni–O hybridization caused by NiO$_6$ octahedral rotations[13] or ionic buckling in the near-surface NiO$_2$ planes[5], but we suggest that charge transfer driven by O vacancies may also play a role.

A close examination of the pre-edge peak shape reveals a shoulder on the higher-energy side of the main peak. Figure 3c shows a Gaussian fitting of the pre-edge peak for the 4 u.c. LaNiO$_3$ film that includes both a main peak at about 527.7 eV and a shoulder peak at about 528.7 eV. The shoulder peak has been observed[24] in LaNiO$_{2.75}$ and LaNiO$_{2.5}$, in which octahedral NiO$_6$ layers alternate with square-planar NiO$_4$ layers[27]. It was attributed to the Ni$^{3+}$ ions occupying the square-planar sites and the higher energy was explained by an increase in the charge-transfer energy due to the lower coordination[24]. In LaNiO$_{3-y}$ with other $y$ values, the intergrowth of the octahedral and square-planar layers is random[26]. If we see the square-planar NiO$_4$ layers as the octahedral NiO$_6$ layers missing apical oxygen, the existence of the shoulder peak in our LaNiO$_3$ films would indicate oxygen vacancies at the apical sites in the films. We further grew LaNiO$_3$ films at lower oxygen pressures of 1 and 5 Pa with all the other growth conditions kept the same. The spectra for three 1.5 u.c. films grown under 1, 5, and 7 Pa oxygen pressures, respectively, are shown in Fig. 3d. The result shows that the shoulder peak becomes more pronounced in the films grown under lower oxygen pressures, confirming its correlation with oxygen vacancies. These results show that even with the high growth oxygen pressure of 7 Pa and the post-growth annealing in 8.5 × 10$^4$ Pa oxygen, oxygen vacancies exist in our LaNiO$_3$ films.

It is difficult to quantify the amount of oxygen vacancies in the LaNiO$_3$ films. However, we can observe the change in the amount of oxygen vacancies with the film thickness by plotting the spectral weight of the shoulder peak, $A_{\text{shoulder}}/A_{\text{total}}$, where $A$ is the area under the peak, as a function of film thickness. The result for the 7 Pa growth oxygen pressure is shown in Fig. 3e. It shows that oxygen vacancies exist in the LaNiO$_3$ films of all the thicknesses at roughly the same level, likely uniform throughout the films. Also, shown in Fig. 3e are the energies of the main and shoulder peaks as functions of the film thickness. The peaks move to higher energies as the film thickness decreases, indicating an upward shift of the conduction band, possibly the consequence of a widening of the band gap, which is consistent with a change toward an insulating state. Since all the films are fully strained to the LaAlO$_3$ substrate and the level of oxygen vacancies is roughly

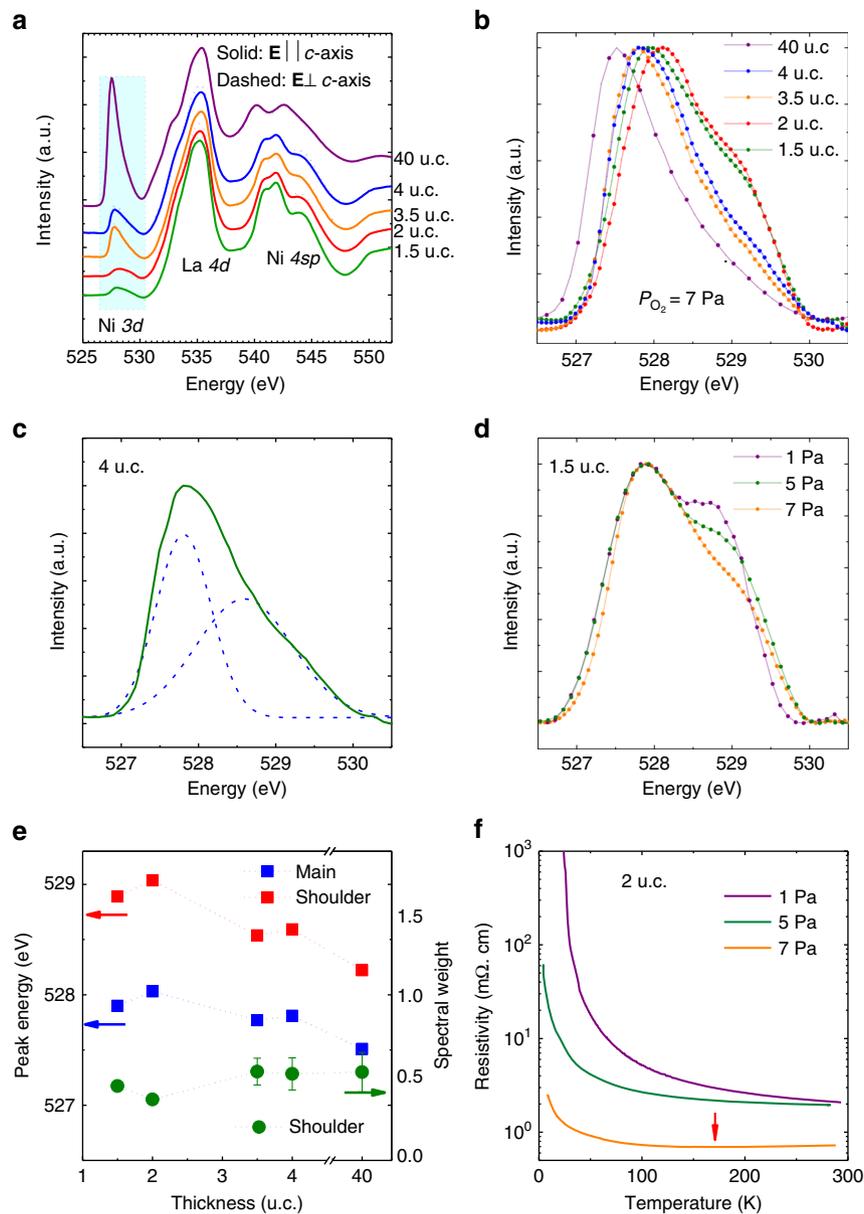

**Fig. 3** X-ray absorption spectroscopy at the oxygen K edge. **a** X-ray absorption spectra at O-K edge for polarizations parallel (solid line) and perpendicular (dashed line) to the c-axis of the LaNiO$_3$ films. **b** Normalized polarization-averaged XAS spectra of the oxygen pre-edge region for films with various thicknesses. **c** Fitting of the pre-edge peak for the 4 u.c. film (green line) using two Gaussian functions (dashed blue line). **d** Normalized pre-edge peak for 1.5 u.c. films grown at 1, 5, and 7 Pa oxygen pressures, respectively. **e** Spectral weight of the shoulder peak, $A_{shoulder}/A_{total}$, where $A$ is the area under the peak, and the energies of the main and shoulder peaks vs. thickness of the LaNiO$_3$ film grown under 7 Pa oxygen pressure. The error bars for the spectral weight are the standard errors of the mean from the Gaussian fitting of the pre-edge peak areas. **f** Temperature-dependent resistivity curves for 2 u.c. LaNiO$_3$ films grown under oxygen pressures of 1, 5, and 7 Pa, respectively. Red arrow points to the metal to insulator transition temperature for the film grown at 7 Pa

the same, the thickness dependence reflects the effect of the dimensional constraint.

The effect of oxygen vacancies on the MIT is shown in Fig. 3f with three 2 u.c. LaNiO$_3$ films grown under different oxygen pressure. A lower growth oxygen pressure leads to a higher resistivity, with the film grown in 7 Pa oxygen being the only one showing metallic behavior at room temperature (the MIT is marked by an arrow). Although we cannot completely rule out the possibility that the film quality varied with oxygen pressure during growth, the results in Fig. 3d, f combined strongly suggests that apical oxygen vacancies are associated with the more insulating behavior. In fact, changes in the valence state and O coordination of Ni ions related to the missing of apical oxygen have been suggested as the mechanism of MIT in reduced LaNiO$_{3-y}$ (ref. [28]). We thus conclude that in addition to the low dimensionality and epitaxial strain that lead to octahedral rotations[13] and ionic buckling[5], oxygen vacancies at the apical sites are an important factor in finally driving the few-unit-cell-thick LaNiO$_3$ films insulating.

**Orbital polarization**. The X-ray linear dichroism (XLD), i.e. the difference between two XAS spectra obtained with different linear polarizations, at the Ni L edge is a sensitive tool for detecting orbital polarization[12,29]. In bulk LaNiO$_3$, the ground state has Ni in the $d^8\underline{L}$ configuration with eight electrons (seven originating

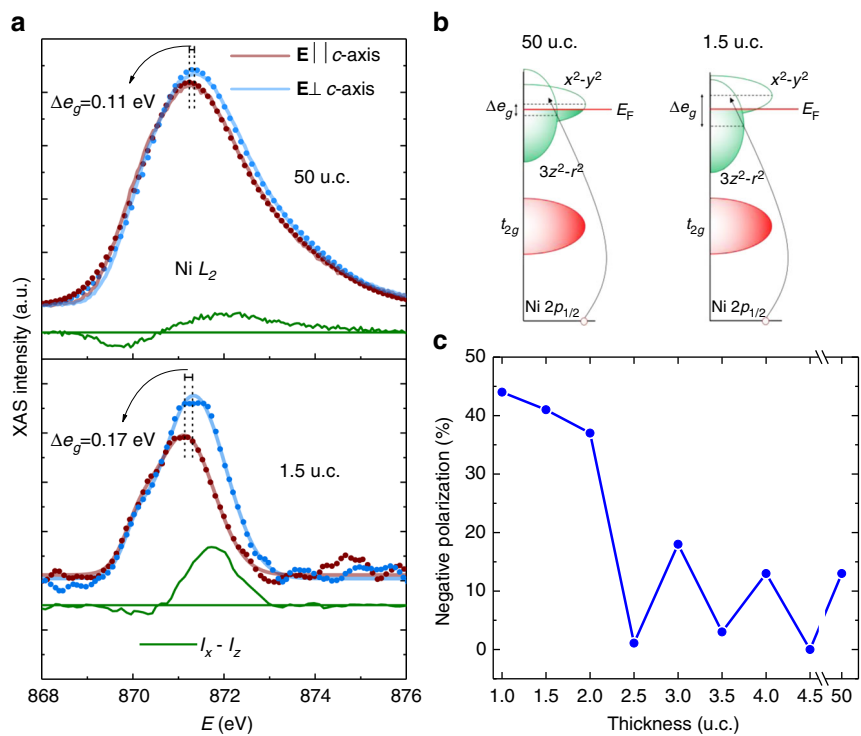

**Fig. 4** Orbital polarization in the LaNiO$_3$ films. **a** XAS spectra at Ni $L_2$-edge for **E**⊥c (blue) and **E**∥c (red) polarizations and X-ray linear dichroism for a 50 u.c. and a 1.5 u.c LaNiO$_3$ films. Solid lines show the Gaussian fit. **b** Schematic of the 3d orbitals for the 50 u.c. and 1.5 u.c LaNiO$_3$ films. $E_F$ denotes the Fermi energy. **c** Magnitude of the negative $d$-orbital polarization vs. film thickness

from the Ni $d$ orbitals and one transferred from the O $p$ orbital), with the $t_{2g}$ orbitals being occupied and the $e_g$ orbitals, $x^2–y^2$ and $3z^2–r^2$, being degenerate[1]. In thin films, epitaxial strain can lift the $e_g$ orbital degeneracy[1,14,16], leading to an orbital polarization P. Linearly polarized X-rays can preferentially probe these strongly directional orbitals. Figure 4a shows normalized polarization-dependent XAS spectra at the Ni $L_2$ edge for the 1.5 u.c. and 50 u.c. LaNiO$_3$ films collected at room temperature. The two spectra correspond to photons polarized parallel and perpendicular to the c-axis of the film, respectively, and the XLD signals are marked by the green areas. For the 1.5 u.c. film, the spectrum for the polarization parallel to the c-axis shifts to lower energies relative to that obtained with perpendicular polarization, indicating a splitting of the $e_g$ orbitals ($\Delta e_g$ in Fig. 4b). For the 50 u.c. film, the shift also occurs but is much smaller. Figure 4c shows P as a function of film thickness, following the approach established by Wu et al.[29] for deriving P from the XLD spectra. The overall negative values of the orbital polarization are consistent with an expansion of the lattice along the direction normal to the film surface under compressive strain[14]. Below 2 u.c. thickness the value for orbital polarization increases substantially to above 40%. For layer thicknesses exceeding 2 u.c., an oscillation of the orbital polarization with the film termination was observed. Evidently, electronic structure and orbital polarization of the LaNiO$_3$ films are directly impacted by the surface termination and the film thickness as also reported by Kumah et al.[5], who cited surface polarization as a reason for such an oscillation. Our DFT + DMFT calculations (Fig. 2g) also show that both the 1 u.c. film and the top layer of the 1.5 u.c. film exhibit negative orbital polarization, although the absolute value of polarization is larger in experiment. Because of the compressive strain of the LaAlO$_3$ substrate and the strongly correlated nature of the $d_{3z^2−r^2}$ orbital, the hybridization of the Ni $d_{x^2−y^2}$ orbital with the O $p$ orbital is increased compared to the Ni $d_{3z^2−r^2}$ –O $p$ hybridization and the hole density of the $d_{x^2−y^2}$ orbital is larger than that of the $d_{3z^2−r^2}$ orbital. The significantly higher orbital polarization in the LaNiO$_3$ films below 2 u.c. may be a reflection of the importance of the surface polarization as depicted in our theoretical results and suggested by Kumah et al.[5]

## Discussion

While our DFT + DMFT calculations predict the LaNiO$_3$ ultrathin films to be metallic whereas experimental observation shows that they are insulating, they do show that the metallic states are fragile and could be easily upset by disorder or oxygen vacancies. The top layer of the 1.5 u.c. film reveals strongly reduced and incoherent spectra at the Fermi energy originated from the oxygen vacancy effect due to the absence of the apical oxygen ions. It is plausible that theoretical calculations of the 1 u.c. ultrathin film with oxygen vacancies would yield an insulating state. It is known that oxygen vacancies affect the electronic structures and transport properties in bulk transition-metal oxides[30,31]. In particular, oxygen vacancies have been shown to induce MIT in LaNiO$_3$[20,24,28]. It was suggested that oxygen vacancies lead to an increase in the Ni$^{2+}$/Ni$^{3+}$ ratio and reduction in bandwidths, thus stronger electron correlation and larger Mott–Hubbard splitting[20,24,28,32]. Our XAS results provide experimental evidence that oxygen vacancies are linked to the MIT in ultrathin LaNiO$_3$ films.

The discovery of oxygen vacancies as a critical factor is important for the fundamental understanding of the MIT in ultrathin LaNiO$_3$ films. Furthermore, oxygen vacancies are an additional parameter to dimensionality and strain for tuning the room-temperature MIT in LaNiO$_3$. For example, electric-field-induced oxygen vacancy formation has been accomplished by electrolyte gating with ionic liquids in VO$_2$[30]. A similar mechanism could turn ultrathin LaNiO$_3$ from showing a metallic to insulating behavior at room temperature, with potential applications in functional devices.

## Methods

**Sample preparation.** ALL-Laser MBE was used to grow the few-unit-cell-thick LaNiO$_3$ films, which also allows us to investigate the properties of the films with different nominal terminations. We found that on our AlO$_2$-terminated LaAlO$_3$ substrate, prepared by HCl etching followed by annealing at 750 °C in flow of oxygen[33], a homoepitaxial LaAlO$_3$ buffer layer is necessary to prepare a high-quality growth template for the ultrathin LaNiO$_3$ films. Ceramic targets of Al$_2$O$_3$, La$_2$O$_3$, and NiO were used. Details for the target preparation can be found in Supplementary Note 1. The ablation was conducted using a KrF excimer laser with 1 Hz repetition rate and 1 J cm$^{-2}$ energy density. Samples were grown at a temperature of 650 °C in oxygen pressure from 1 to 7 Pa. After the deposition of each LaO or NiO$_2$ layer, a 100-s dwell time was used to oxygenate the film. Most films presented in this work were grown under 7 Pa of oxygen, the highest oxygen pressure we could use. We used RHEED intensity oscillation to control growth, and above 7 Pa the diffraction spots were no longer sharp. To further oxygenate the films, post-growth annealing in $8.5 \times 10^4$ Pa of oxygen was performed at the growth temperature of 650 °C for 30 min. See Supplementary Notes 2 and 3 for homoepitaxial LaAlO$_3$ buffer layer and LaNiO$_3$ layer growth and characterization.

**Transport measurements.** In-plane transport measurements were carried out between 4.2 K and room temperature by using the van der Pauw configuration with contacts at the corners of the films with 5 mm × 5 mm surface area.

**X-ray absorption spectroscopy.** X-ray absorption and linear dichroism measurements were carried out at Beamline 4.0.2 of the Advanced Light Source at Lawrence Berkeley National Laboratory and Beamline U4B of the National Synchrotron Light Source at Brookhaven National Laboratory in TEY mode with an energy resolution of ~0.1 eV. The average probing depth for soft X-rays in the TEY mode was estimated to be about 5 nm. The X-ray spot was focused down to ~100-μm diameter in the normal incidence geometry. Samples were characterized using nearly 100% horizontally or vertically polarized X-ray beams at an incidence angle of 15° from the sample plane. In such a grazing measurement geometry the X-ray electric field is oriented parallel to the surface for vertically polarized light, and almost perpendicular to the surface for horizontally polarized light, thus providing maximum sensitivity to changes in orbital character along the different crystallographic axes.

**XRR measurement.** The XRR measurement was conducted at Beamline X14A of the National Synchrotron Light Source at Brookhaven National Laboratory. The X-ray wavelength was 0.7793 Å. The incident beam size was 0.3 mm (v) x 1.5 mm (h) and the reflectivity was measured using an Si avalanche photodiode detector with a Ge (111) crystal analyzer. The simulation of the reflectivity curves was performed with the IMD code[34] available in the XOP package. The layer densities were 7.2 g cm$^{-3}$ for the LaNiO$_3$ layer and between 5.8 g cm$^{-3}$ for the LaAlO$_3$ buffer layer.

**STEM measurements.** TEM samples were prepared by adhering the as-grown LaNiO$_3$ film to a sacrificial Si crystal with M-Bond adhesive. Cross-sectional cuts were made through this stack, and the sample was thinned by mechanical polishing in a wedge geometry followed by final thinning using a Gatan PIPS-II system. All data were acquired on the Titan-class TEAM 0.5 instrument at the Molecular Foundry at Lawrence Berkeley National Laboratory with geometric aberrations in the probe forming optics corrected to the third order. Data were acquired at an accelerating voltage of 300 kV with a 17 mrad convergence semiangle. The annular bright field collection semiangle was 6–25 mrad and the simultaneously acquired annular dark field images had a collection semiangle of 53–270 mrad. Images were acquired in sequential pairs with a 90° rotation offset in order to correct for the linear and nonlinear instrumental drift according to Ophus et al.[18]. The LaNiO$_3$ lattice parameter was measured by fitting 2D Gaussian functions to the La lattice sites from a mean slab perpendicular to the surface averaged over 20 u.c. parallel to the surface for an improved signal to noise ratio. The bulk LaAlO$_3$ 10 u.c. from the surface were used as in internal calibration reference.

**First-principles calculation.** We adopt DFT for the structural relaxation and both DFT + DMFT and DFT for the DOS calculation in both 1 u.c and 1.5 u.c. LaNiO$_3$ thin films on the LaAlO$_3$ substrate and bulk LaNiO$_3$. We use the Vienna Ab-initio Simulation Package (VASP) code for the DFT calculations by taking the exchange-correlation potential to be the generalized gradient approximation using the Perdew–Burke–Ernzerhof (PBE) functional. A k-point mesh of 6 × 6 × 1 is used for the slab calculation while the mesh of 6 × 6 × 6 is used for bulk. The plane-wave energy cutoff $E_{cut}$ is set to be 600 eV. All internal atomic coordinates and the (001) axis lattice parameter of both structures are fully relaxed for both structures while the in-plane lattice constants are fixed. The tetragonal symmetry is imposed during the structural relaxation meaning that the in-plane lattice constants are fixed to the LaAlO$_3$ in-plane constant (~3.81 Å) while the angles between lattice vectors are constrained to be at right angles. Both tilting and rotation of NiO$_6$ octahedra are also allowed as each Ni–O layer contains two Ni ions. We adopt a symmetrical geometry of the slab, namely the vacuum/La-O/Ni-O/LaAlO$_3$/Ni-O/La-O/vacuum geometry for the 1 u.c LaNiO$_3$ film with the La-O layer termination and the vacuum/Ni-O/La-O/Ni-O/ LaAlO$_3$/Ni-O/La-O/Ni-O/vacuum geometry for the 1.5 u.c. LaNiO$_3$ film to avoid the expensive simulation of semi-infinite LaAlO$_3$ layers. The thickness of the LaAlO$_3$ substrate used in both the 1 u.c and 1.5 u.c LaNiO$_3$ film calculations was 6 u.c. for this symmetrical geometry of the slab. The thickness of the vacuum is as large as 6 Å. We have checked that increasing the vacuum thickness to 10 Å does not change the physics discussed in the paper. Namely, it produces a similar DOSs as shown in Fig. 2d.

For DFT + DMFT calculations, we first construct the localized subspace of Ni $d$ and O $p$ orbitals from the DFT band structure spanning the 12 eV energy range of the Ni $d$ and O $p$ band complex by using the maximally localized Wannier function, then perform the DMFT self-consistent calculations for the correlated subspace of Ni $d$ orbitals. We used the continuous time quantum Monte Carlo method[35,36] to solve the impurity problem of DMFT. For interaction parameters, we chose the Hubbard interaction $U = 5$ eV and the Hund's coupling $J = 1$ eV.

**Data availability.** The data that support the findings of this study are available from the corresponding author upon reasonable request.

### Acknowledgements
Work at Temple University was supported by NSF under Grant No. DMR-1245000 (to M.G. and X.X.X.). A.X.G and R.U.C. acknowledge support from the U.S. Army Research Office, under Grant No. W911NF-15-1-0181. The National Synchrotron Light Source at Brookhaven National Lab was supported by the Office of Science, Office of Basic Energy Sciences, of the US Department of Energy under Contract No. DE-AC02-98CH10886. The Advanced Light Source is supported by the Director, Office of Science, Office of Basic Energy Sciences, of the U.S. Department of Energy under Contract No. DE-AC02-05CH11231. The authors thank Marissa Libbee for assistance with the TEM sample preparation.

### Author contributions
X.X.X. conceived and designed the experiments, and M.G. and Q.Y.L. carried out the development of the ALL-Laser MBE technique. M.G. fabricated the samples. M.G. and L.K. performed the XRD and transport measurements. A.X.G. designed and supervised the spectroscopic measurements and analyses. R.U.C., G.E.S., D.A.A., P.S., P.O., and E.A. performed the spectroscopic measurements. H.P. and A.J.M. performed the DFT calculations. M.G. and B.A.D. analyzed the spectroscopy data. J.M.B. performed the XRR measurement. P.O. and J.M.B. analyzed the XRR data. J.C. performed the TEM measurements. All authors contributed to the writing of the manuscript.

### Additional information
**Supplementary Information** accompanies this paper at https://doi.org/10.1038/s41467-018-04546-5.

**Competing interests:** The authors declare no competing interests.

**Reprints and permission** information is available online at http://npg.nature.com/reprintsandpermissions/

**Publisher's note:** Springer Nature remains neutral with regard to jurisdictional claims in published maps and institutional affiliations.